\documentclass[aps,pre,groupedaddress]{revtex4-2}

\usepackage{siunitx} 
\usepackage{amsmath} 
\usepackage{graphicx} 
\usepackage{subcaption} 
\usepackage{xcolor} 

\newcommand{\new}[1]{\textcolor{black}{#1}}

\definecolor{linkcolor}{rgb}{0,0,0.6} 
\usepackage[pdftex,colorlinks=true, pdfstartview=FitV, linkcolor= linkcolor, citecolor= linkcolor, urlcolor= linkcolor, hyperindex=true,hyperfigures=true]{hyperref} 

\begin{document}


\title{Comment on ``Harvesting information to control non-equilibrium states of active matter''}


\author{Antoine B\'{e}rut}
\email[]{antoine.berut@univ-lyon1.fr}
\affiliation{Universit\'{e} de Lyon, Universit\'{e} Claude Bernard Lyon 1, CNRS, Institut Lumi\`{e}re Mati\`{e}re, F-69622, Villeurbanne, France}


\date{\today}

\begin{abstract}
In the article \textit{Phys. Rev. E} \textbf{106}, 054617 ``Harvesting information to control non-equilibrium states of active matter'', the authors study the transition from one non-equilibrium steady-state (NESS) to another NESS by changing the correlated noise that is driving a Brownian particle held in an optical trap. They find that the amount of heat that is released during the transition is directly proportional to the difference of spectral entropy between the two colored noises,  in a fashion that is reminiscent of Landauer's principle. In this comment I argue that the found relation between the released heat and the spectral entropy does not hold in general, \new{and that}  one can provide examples of noises where it clearly fails. I also show that, even in the case considered by the authors, the relation \new{cannot be} rigorously true and is only approximately verified experimentally.
\end{abstract}


\maketitle

\section*{Introduction}

In the article ``Harvesting information to control non-equilibrium states of active matter''~\cite{Goerlich2022} the authors consider a Brownian particle, held in an optical trap, and submitted to an external correlated (\textit{i.e.} ``colored'') noise $\eta$. The displacement of the particle is accurately described by the overdamped Langevin equation:
\begin{equation}
\label{eq:Langevin}
\dot{x} = -\frac{\kappa}{\gamma} x + \sqrt{2D} \xi + \sqrt{2D_a} \eta
\end{equation}
Where $\kappa$ is the trap stiffness, $\gamma$ is the Stokes viscous drag ($\gamma = 6\pi R \mu$ with $R$ the particle's radius, and $\mu$ the fluid's viscosity), $\xi$ is a Gaussian white noise with unit variance, accounting for the thermal fluctuations of the bath, $D$ is the equilibrium diffusion coefficient ($D=k_\mathrm{B}T/\gamma$ with $k_\mathrm{B}$ the Bolztmann constant and $T$ the temperature), $D_a$ is a control parameter that allows to change the amplitude of the colored noise $\eta$.\bigskip

The authors then consider what they call a ``STEP Protocol'': at time $t_1$ the particle is initially in a non-equilibrium steady-state (NESS) with a given \new{colored} noise $\eta_1$, then at a given time $t_c > t_1$ the noise is abruptly changed to another colored noise $\eta_2$, and the particle is let to reach a different NESS at time $t_2 > t_c$. Using the framework of stochastic thermodynamics~\cite{Sekimoto1998,Seifert2012}, they compute the average amount of \textit{excess} heat that is released in the transition between the two NESS: $\Delta Q_{EX}$.Their main result, is then that this amount of heat is directly proportional to the difference of spectral entropy (which is the Shannon entropy in the frequency domain~\cite{Zaccarelli2013}), $\Delta H_S$ between the two noises $\eta_1$ and $\eta_2$:
\begin{equation}
\label{eq:main_result}
\Delta Q_{EX} \propto \Delta H_S
\end{equation}
They claim that this relation is ``akin to Landauer's principle'', as it relates a thermodynamics quantity ($\Delta Q_{EX}$), to an informational quantity ($\Delta H_S$). Their interpretation of the result is that ``the protocol harvests information from the colored noise, turns it into heat necessary for the transition between the two NESS, and finally releases it to the surrounding environment''.\bigskip

In this comment, I argue that the found relation between the heat released and spectral entropy is not true in general, and that, even in the exact case considered by the authors, it does not hold rigorously. In a first section~\ref{sec:theory}, \new{starting from the definitions of stochastic heat and spectral entropy, I derive a general equation (\ref{eq:general_case}) to compute both quantities for any kind of colored noise. This result shows that the claimed relation~\ref{eq:main_result} is invalid for the vast majority of noises.} \new{In a second section \ref{sec:examples} I  give} several examples of noises for which the relation clearly fails. In the third section~\ref{sec:authors_case}, I discuss analytical results obtained \new{when one considers the exact situation described} by the authors, and show numerical simulations \new{to verify that, even in this case, the relation~\ref{eq:main_result} fails}. Finally, I discuss a possible explanation to why the relation \new{seems} approximately verified in the experimental data of the article~\citep{Goerlich2022}.

\new{\section{Theoretical calculations\label{sec:theory}}}

We first consider the two quantities of interest $\Delta Q_{EX}$ and $\Delta H_S$.\bigskip

The spectral entropy $H_S$ of a noise $\eta$ is defined as the Shannon entropy applied to the Power Spectral Density (PSD) of that noise~\cite{Zaccarelli2013}:
\begin{equation}
\label{eq:spectral_entropy}
H_S(\eta) = \sum_{j=1}^{N} S_{\eta}^*(\omega_j)\ln \left(S_{\eta}^*(\omega_j) \right)
\end{equation}
with $S_{\eta}^*(\omega_j)$ the normalized PSD of the noise at the (discretized) angular frequency $\omega_j$:
\begin{equation}
S_{\eta}^*(\omega_j) = S_\eta(\omega_j) / \sum_{j=1}^{N} S_\eta(\omega_j)
\end{equation}
where $S_\eta$ stands for the PSD of the noise $\eta$.\\
This quantity is a measurement of the ``flatness'' of the noise's PSD. As stated by the authors: it vanishes for a monochromatic signal, it reaches its maximum $\ln(N)$ for a white noise, and any correlated noise has an intermediate value of $H_S$.\bigskip

The average amount of \textit{excess} heat that is released in the transition between two NESS is given by~\cite{Sekimoto1998,Seifert2012}:
\begin{equation}
\label{eq:released_heat}
\Delta Q_{EX} = \frac{1}{2} \int_{t_1}^{t_2} \kappa \frac{\mathrm{d}\langle x^2 \rangle}{\mathrm{d}t} \mathrm{d}t^\prime 
\end{equation}
where $\kappa$ is the trap stiffness, $t_i$ (with $i\in \{1,2\}$) is a time at which the particle is in a NESS with the colored noise $\eta_i$, and $\langle \cdot \rangle$ stands for the ensemble average. This relation can be written:
\begin{equation}
\Delta Q_{EX} = \frac{\kappa}{2} \left( \sigma^2_{x}(t_2) - \sigma^2_{x}(t_1)  \right)
\end{equation}
where $\sigma^2_x$ stands for the variance of the particle's position ($\sigma^2_x = \langle x^2 \rangle$).\\
Thus, we see that the amount of heat dissipated when the external noise is switch from $\eta_1$ to  $\eta_2$ can simply be calculated by measuring the variance of the particle's position in the first NESS $\sigma^2_{x_1}$ and in the second NESS $\sigma^2_{x_2}$. \new{In the article~\cite{Goerlich2022}, the authors have only considered those equations in the case of an exponentially-correlated noise, and state that ``[the] extension to other classes of noises is less clear''. In the following, we show that it is actually possible to analytically compute both $\Delta H_S$ and $\Delta Q_{EX}$ \emph{for any kind of colored noise}, as long as their PSD is known.}\bigskip

To \new{do that}, we recall that in a NESS the system is both stationary and ergodic~\footnote{Both stationarity and ergodicity are required so that the ensemble average and time average are equals.}. Therefore, the Wiener–Khinchin theorem~\cite{Wiener1930,Khintchine1934} allows us to link the variance of the position to the PSD of the particle's position $S_x$, through the relation:
\begin{equation}
\begin{split}
\label{eq:variance}
\sigma^2_{x_i} & = \frac{1}{2\pi} \int_{-\infty}^\infty S_{x_i}(\omega) \, \mathrm{d}\omega \\
 & = \frac{1}{2\pi} \int_{-\infty}^\infty \frac{1}{\omega^2 + \omega_0^2} \left( 2D S_{\xi}(\omega) + 2D_a S_{\eta_i}(\omega) \right) \, \mathrm{d}\omega
\end{split}
\end{equation}
where $\omega_0$ is the natural frequency of the system ($\omega_0 = \kappa / \gamma$), $S_{\xi}$ is the PSD of the Gaussian white noise (which is a constant), and $S_{\eta_i}$ is the PSD of the noise $\eta_i$.\\
In the end, for a STEP protocol where the noise is switch from $\eta_1$ to $\eta_2$, we have the formulas:
\begin{equation}
\label{eq:general_case}
\begin{split}
\Delta H_S & = \sum_{j=1}^{N} \left[ S_{\eta_1}^*(\omega_j)\ln \left(S_{\eta_1}^*(\omega_j) \right) - S_{\eta_2}^*(\omega_j)\ln \left(S_{\eta_2}^*(\omega_j) \right) \right] \\
\Delta Q_{EX} & = \frac{\kappa}{2\pi} \int_{-\infty}^\infty \frac{2D_a}{\omega^2 + \omega_0^2} \left[ S_{\eta_2}(\omega) - S_{\eta_1}(\omega) \right] \, \mathrm{d}\omega
\end{split}
\end{equation}\bigskip

As one can see, both quantities $\Delta Q_{EX}$ and $\Delta H_S$ are directly linked to the PSD of the two noises $S_{\eta_1}$ and $S_{\eta_2}$. However, there are critical differences:
\begin{itemize}
\item $H_S(\eta)$ only depends on the ``flatness'' of $S_{\eta}$, but it does not depend on the noise amplitude, nor on the way that the power is distributed along frequencies. Since the terms of the sum~\ref{eq:spectral_entropy} can be permuted without changing the result, a PSD that has ``most of its power at high frequencies'' can have the same spectral entropy as a PSD that has ``most of its power at low frequencies''.
\item On the contrary, the position variance $\sigma^2_x$, depends explicitly on the noise amplitude, and the frequency content of the PSD. Indeed, as seen in equation~\ref{eq:variance}, $\sigma^2_x$ is the integral of $S_\eta(\omega)$ weighted by the mechanical response of the system (the term $1/(\omega^2 + \omega_0^2)$ here). Since the overdamped Brownian particle trapped in optical tweezers acts as a first-order low-pass filter with cut-off frequency $\omega_0$, only the ``low frequency content'' of the noise's PSD will have a significant impact on the variance of particle's position.
\end{itemize}

\new{We want to stress that equation~\ref{eq:general_case} alone contradicts the main result of the article~\cite{Goerlich2022}. For the very vast majority of usual colored noises, the relation of direct proportionality between the spectral entropy difference $\Delta H_S$ and the released heat $\Delta Q_{EX}$ is not verified. In the following section~\ref{sec:examples}, as a pedagogical example, we provide several couples  $\eta_1$ and $\eta_2$ that are designed} such that they have the same spectral entropy ($\Delta H_S = 0$), but still produce a non-zero amount of heat released when the STEP protocol is applied ($\Delta Q_{EX} \neq 0$), which \new{clearly shows the failure of the relation~\ref{eq:main_result}}.

\section{Examples of noises for which the spectral entropy is not directly proportional to the heat released\label{sec:examples}}

We first note that, in the article~\citep{Goerlich2022}, the authors have considered noises with the same amplitude to study only the influence of the noise's ``color'' on $\Delta H_S$ and $\Delta Q_{EX}$. Therefore, in the rest of this comment, we will only consider noises that are normalized so that their variance is equal to one.\bigskip

\begin{figure}[!ht]
\begin{subfigure}{.45\textwidth}
  \centering
  \includegraphics[width=\linewidth]{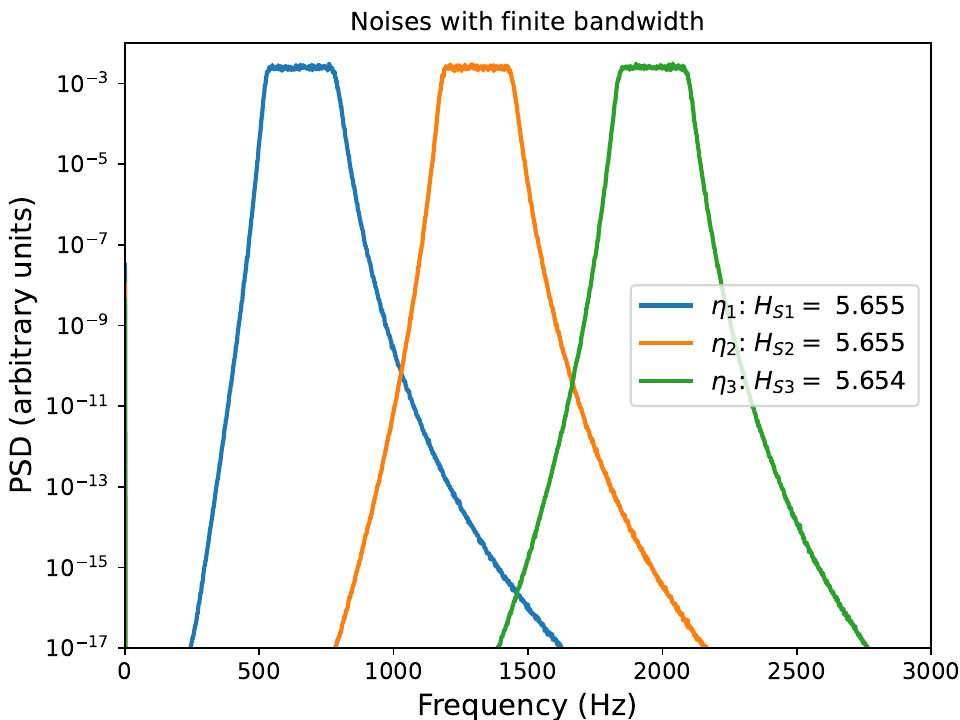}
  \caption{\label{fig:1a}}  
\end{subfigure}%
\begin{subfigure}{.45\textwidth}
  \centering
  \includegraphics[width=\linewidth]{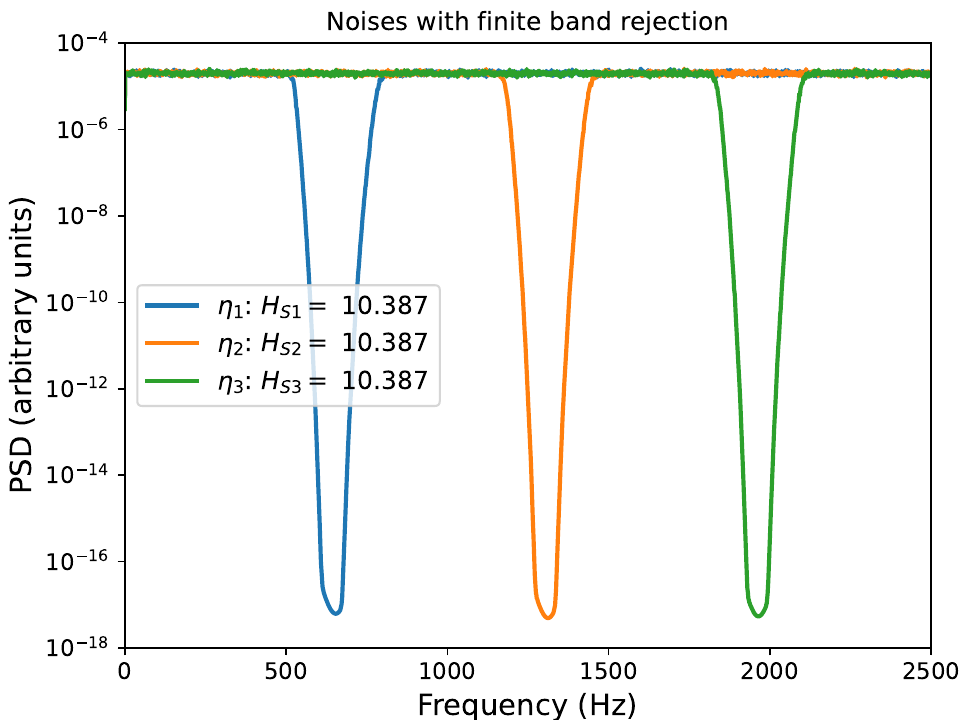}
  \caption{\label{fig:1b}}
\end{subfigure}
\caption{Examples of noises with a finite bandwidth (a) or finite band rejection (b), centered around different frequencies. As shown in legend, all the noises in each category have the same spectral entropy $H_S$ regardless of the value of the central frequency. Noises are generated by numerically filtering a Gaussian white noise, the spectral entropy is computed using equation~\ref{eq:spectral_entropy}.}
\label{fig:fig1}
\end{figure}

The simplest way to design noises that have the same spectral entropy, but do not result in the same variance for the particle's position, is to consider noises with a finite bandwidth, or noises with a finite band rejection in their PSD. Examples are shown in figure~\ref{fig:fig1}. In this case, the value of the spectral entropy $H_S$ is the same for each noise with the same finite bandwidth (respectively band rejection), regardless of the central frequency of the band-pass (respectively band-stop) range. On the contrary, the value of the particle's position variance $\sigma^2_x$, highly depends on this central frequency. Indeed, we recall that the overdamped trapped particle acts as a low-pass filter: $\sigma^2_x$ will be higher for a noise with a high power at low frequency (such as the blue curve $\eta_1$ in figure~\ref{fig:1a}) than for a noise with a high power at high frequency (such as the green curve $\eta_3$ in figure~\ref{fig:1a}). Therefore, using such noises, it is possible to design a STEP protocol with no change of spectral entropy ($\Delta H_S = 0$) while having a non-zero amount of heat released ($\Delta Q_{EX} \neq 0$).\bigskip

\begin{figure}[!ht]
\begin{subfigure}{.45\textwidth}
  \centering
  \includegraphics[width=\linewidth]{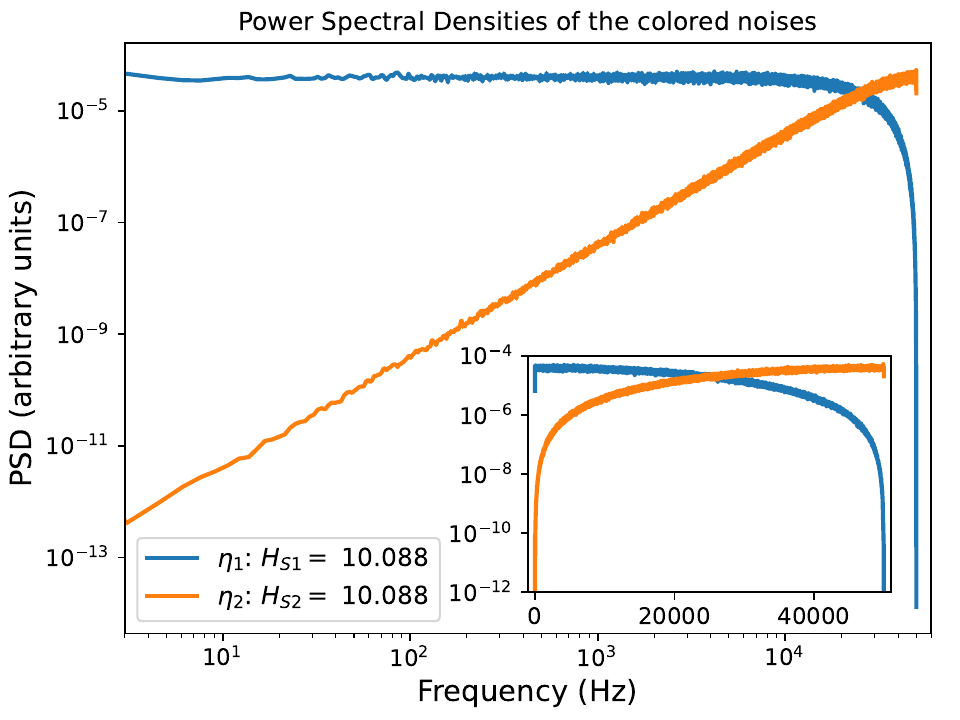}
  \caption{\label{fig:2a}}
\end{subfigure}%
\begin{subfigure}{.45\textwidth}
  \centering
  \includegraphics[width=\linewidth]{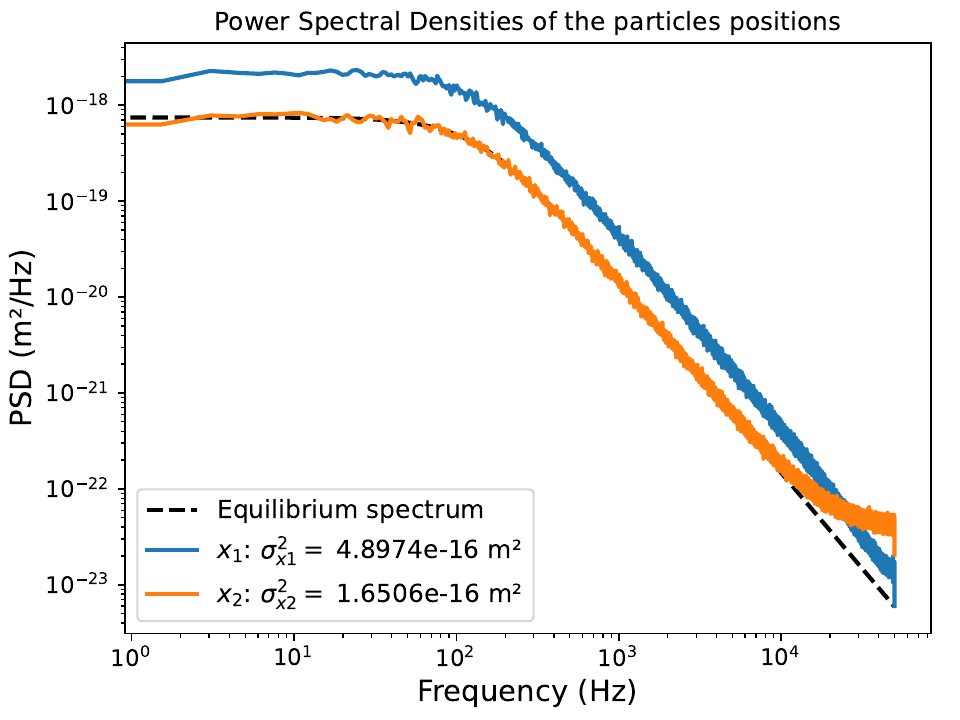}
    \caption{\label{fig:2b}}
\end{subfigure}
\caption{Example of two continuous noises with the same variance and same spectral entropy that do not result in the same variance for the Brownian particle to which they are submitted. (a) PSD of two noises $\eta_1$ (blue) and $\eta_2$ (orange) designed to have the same spectral entropy $H_S$ (values indicated in legend). Inset: same curves plotted in \textit{semilogy} to show that the cut-off frequencies are indeed chosen such that the two PSD crosses in the middle of the considered frequency range. The noises are obtained by numerically filtering a Gaussian white noise. (b) PSD of the Brownian particle's position when it is submitted to no external noise (black-dashed), to $\eta_1$ (blue), and to $\eta_2$ (orange) respectively. PSD clearly show different behaviors (values of variances $\sigma^2_{x1}$ and $\sigma^2_{x2}$ indicated are in legend). The particle position is obtained by numerically integrating the Langevin equation~\ref{eq:Langevin} using Heun's method~\citep{Mannella2002}.}
\label{fig:fig2}
\end{figure}

Of course, we are not limited to noises with a finite bandwidth or finite band rejection. It is also possible to design noises with a continuous PSD that have the same spectral entropy, but do not result in the same variance for the particle's position. For example, one can consider a noise $\eta_1$ with a low-frequency content, and a noise $\eta_2$ with a high frequency content:
\begin{equation}
S_{\eta_1}(\omega) \propto \frac{1}{\omega_{c1}^2+\omega^2} \quad \mathrm{and} \quad S_{\eta_2}(\omega) \propto \frac{\omega^2}{\omega_{c2}^2+\omega^2}
\end{equation}
If the two cut-off frequencies $\omega_{c1}$ and $\omega_{c2}$ are chosen such that the two PSD of the noises crosses exactly in the middle of the considered frequency range, they will have the same spectral entropy ($\Delta H_S =0$). However, it is clear that the two noises will not produce the same position variance when they are applied to the particle. For example, suppose that the two cut-off frequencies are bigger than the natural frequency of the particle $\omega_{c1} \gg \omega_0$ and $\omega_{c2} \gg \omega_0$. Then, $\eta_1$ will act nearly as a white noise on the particle, and on the contrary, $\eta_2$ will have nearly no influence on it's position. This will result in  $\sigma^2_{x1} > \sigma^2_{x2}$, and therefore $\Delta Q_{EX} \neq 0$.\\
An example is shown in figure~\ref{fig:fig2} for $\omega_{c1} = \omega_{c2} = \SI{157079}{\radian\per\second}$ (corresponding to \SI{25000}{\hertz}), and  $\omega_0 = \SI{884}{\radian\per\second}$ (corresponding to \SI{140}{\hertz}). As seen in figure~\ref{fig:2a}, the two noises $\eta_1$ and $\eta_2$ have the same variance and the same spectral entropy. However, as seen in figure~\ref{fig:2b}, the PSD of the particle's position $S_{x1}$ and $S_{x2}$, obtained by numerically integrating the Langevin equation~\ref{eq:Langevin} with $\eta_1$ and $\eta_2$ respectively, are very different, and do not produce the same variance for the particle's position.\bigskip

Thus, we have seen that it is possible to design noises such that the relation of direct proportionality between $\Delta Q_{EX}$ and $\Delta H_S$ clearly fails in the STEP protocol~\footnote{Note that it is also possible to design noises such that they do not have the same spectral entropy ($\Delta H_S \neq 0$), but produce the same variance of the particle's position ($\Delta Q_{EX} = 0$).}. This \new{further emphasizes} that the relation is not true in general, and cannot be used with any colored noise. In the next section we will show that, even in the exact case that is considered by the authors in the article~\citep{Goerlich2022}, the relation is \new{does not hold rigorously}.

\section{The spectral entropy is not directly proportional to the heat released even in the case considered by the authors\label{sec:authors_case}}

\subsection{Analytical and simulated results of the case considered by the authors}

In the article~\citep{Goerlich2022}, the colored noise $\eta$, that is driving the Brownian particle, is generated by an Ornstein-Uhlenbeck process:
\begin{equation}
\label{eq:colored_noise}
\mathrm{d}\eta = -\omega_c \eta \mathrm{d}t + \sqrt{2\alpha\omega_c} \mathrm{d}W
\end{equation}
where $\mathrm{d}W$ is a $\delta$-correlated Wiener process. This noise $\eta$ is characterized by its variance $\alpha$, and by its cut-off angular frequency $\omega_c$. It's Power Spectral Density (PSD) is given by:
\begin{equation}
\label{eq:PSD_colored_noise}
S_{\eta}(\omega) = \frac{2 \alpha \omega_{c}}{\omega_{c}^2+\omega^2} \\
\end{equation}
For the STEP protocol, the authors have chosen to keep the amplitude of the colored noise constant ($\alpha_1 = 1 = \alpha_2$), and to only change its cut-off frequency from $\omega_{c1}$ to $\omega_{c2}$. \new{Then, using the PSD of the noises~\ref{eq:PSD_colored_noise}, and the general formulas~\ref{eq:general_case}, we directly obtain}:
\begin{equation}
\label{eq:analytical_results}
\begin{split}
\Delta H_S & = \ln \left( \frac{\omega_{c1}}{\omega_{c2}} \right) \\
\Delta Q_{EX} & = \frac{\kappa D_a}{\omega_0} \frac{(\omega_{c1}-\omega_{c2})}{(\omega_0+\omega_{c2})(\omega_0+\omega_{c1})}
\end{split}
\end{equation}
Therefore, we see that the relation $\Delta H_S \propto \Delta Q_{EX}$ fails analytically, even in the case considered by the authors. \new{Note that those results are identical to the ones obtained by the authors, namely eqs. (H5) and (G9) in the Sup. Mat. of~\cite{Goerlich2022} (except for a missing $\kappa$ prefactor in equation (G9) that is probably a typo).}\bigskip

\begin{figure}[!ht]
\begin{subfigure}{.45\textwidth}
  \centering
  \includegraphics[width=\linewidth]{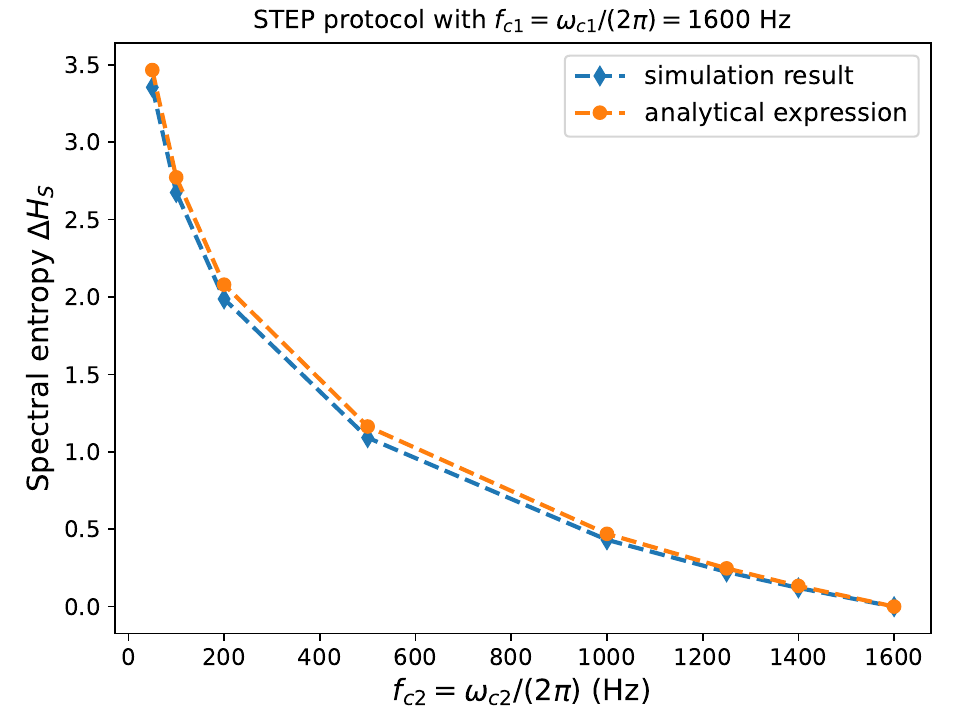}
  \caption{\label{fig:3a}}
\end{subfigure}%
\begin{subfigure}{.45\textwidth}
  \centering
  \includegraphics[width=\linewidth]{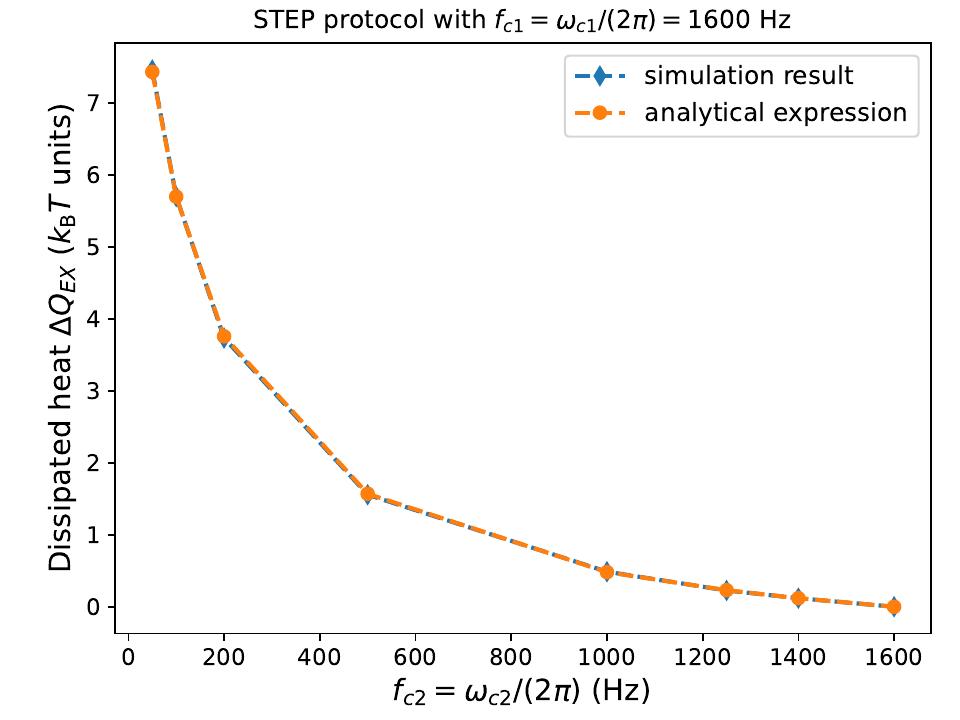}
    \caption{\label{fig:3b}}
\end{subfigure}
\caption{Spectral entropy difference $\Delta H_S$ (a) and dissipated heat $\Delta Q_{EX}$ (b) obtained from numerical simulations of a STEP protocol where the colored noise is changed from $\eta_1$ ($\alpha_1 = 1$ and $f_{c1} = \SI{1600}{Hz}$) to $\eta_2$ ($\alpha_2 = 1$ and $f_{c2} \in [50,100,200,500,1000,1250,1400,1600] \, \si{\hertz}$). The simulated results are compared to the analytical expressions~\ref{eq:analytical_results}.}
\label{fig:fig3}
\end{figure}

\new{The failure of the linearity relation} can be verified in numerical simulations. Using Heun's scheme~\cite{Mannella2002}, we have first integrated equation~\ref{eq:colored_noise} to obtain colored noises $\eta_i$ with the desired properties. Then, we have integrated the Langevin equation~\ref{eq:Langevin} during a STEP protocol, where we change the noise from $\eta_1$ to $\eta_2$ (characterized by $\alpha_1 = 1 = \alpha_2$ and $\omega_{c1} \neq \omega_{c2}$) in the middle of the trajectory. We have used parameters close to the experimental ones~\footnote{Note that the exact values of the parameters are not important here, as the numerical results are only presented to ``verify'' the analytical calculations, which are expected to hold for any set of parameters.}: bead radius $R = \SI{1.5}{\micro\meter}$, water viscosity $\mu = \SI{1e-3}{\pascal\second}$, temperature $T = \SI{298.15}{\kelvin}$, trap stiffness $\kappa = \SI{25}{\pico\newton/\micro\meter}$, Boltzmann constant $k_\mathrm{B} = \SI{1.381e-23}{\joule/\kelvin}$, amplification of the noise $D_a = 10000 D$ with $D=k_\mathrm{B}T/\kappa \approx \SI{1.456e-13}{\square\meter/\second}$, cut-off frequency of first noise $f_{c1} = \omega_{c1}/(2\pi) = \SI{1600}{\hertz}$, cut-off frequency of second noise $f_{c2} = \omega_{c2}/(2\pi) \in [50,100,200,500,1000,1250,1400,1600] \, \si{\hertz}$, compared to $f_0 = \omega_0/(2\pi) \approx \SI{140}{\hertz}$, and integration time-step $\delta t = \SI{1e-5}{\second}$. Finally, we have computed both $\Delta H_S$ and $\Delta Q_{EX}$ for a thousand runs, and compared the mean results to the predictions~\ref{eq:analytical_results}. \new{All the numerical simulation Python codes are available on a Zenodo depository~\cite{codeZenodo}, and can be consulted directly online on the associated GitHub page~\cite{GithubLink}}.\bigskip

As seen in figure~\ref{fig:fig3}, both quantities correctly follow the analytical predictions~\ref{eq:analytical_results}, within the numerical accuracy. \new{We stress that} there is no reason to believe that these results would not hold experimentally, unless the Langevin equation~\ref{eq:Langevin} does not correctly describe the experimental system\new{~\footnote{\new{Note that if the Langevin equation does not accurately describe the system, the formulas that are derived from this equation are not valid anymore. In particular the definition of the stochastic heat~\ref{eq:released_heat}, that is used by the authors in their article, does not hold in this case.}}}.

\subsection{Why the relation seems verified experimentally?}

\new{So far, we have seen that the relation of direct proportionality between the spectral entropy difference and the heat released~\ref{eq:main_result} does not hold in the general case. We have also seen that it is not verified analytically, nor numerically, in the case considered by the authors in~\cite{Goerlich2022}. Therefore, one can wonder how this relation can appear to be experimentally verified. In this subsection, we propose an explanation, based on the way that was used to verify the validity of the relation. We show that the graphical representation chosen by the authors can lead to wrong interpretations, and propose a more reliable way to plot the data.}\bigskip 

Experimentally, the authors have found a good agreement for the relation:
\begin{equation}
\frac{\Delta Q_{EX}}{k_\mathrm{B}T_{eq}} = \Delta H_S
\end{equation}
where $T_{eq}$ is an experimental quantity, equal to the effective temperature ($T_{eq}=\kappa \langle x^2 \rangle/k_\mathrm{B}$) that is measured when the particle is submitted to a white noise with the same amplitude as the colored noise.\bigskip

\begin{figure}[!ht]
\begin{subfigure}{.45\textwidth}
  \centering
  \includegraphics[width=\linewidth]{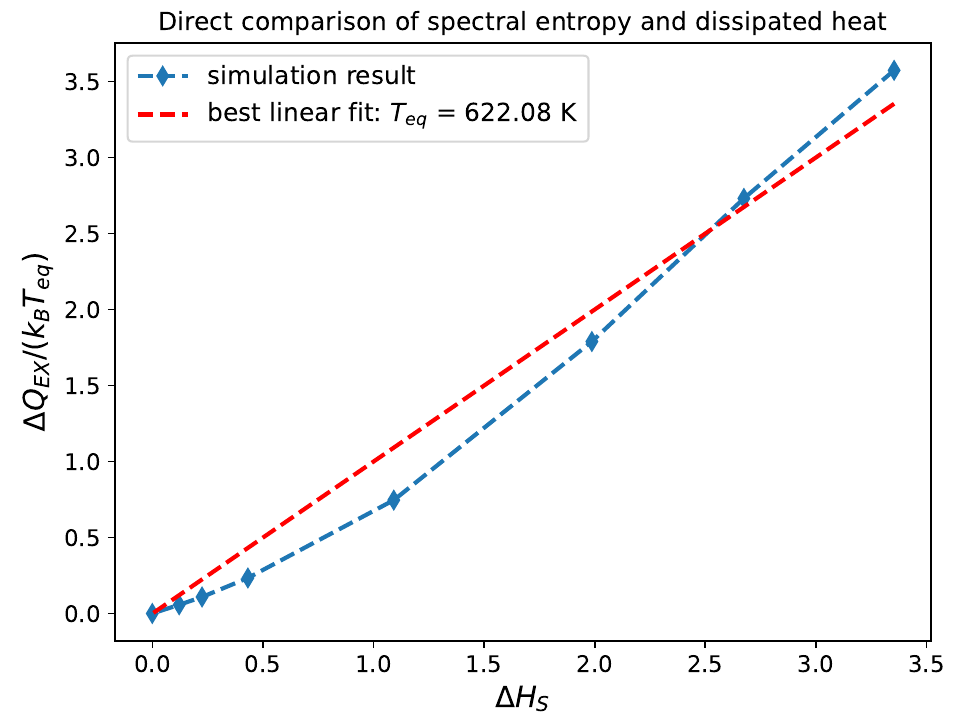}
  \caption{\label{fig:4a}}
\end{subfigure}%
\begin{subfigure}{.45\textwidth}
  \centering
  \includegraphics[width=\linewidth]{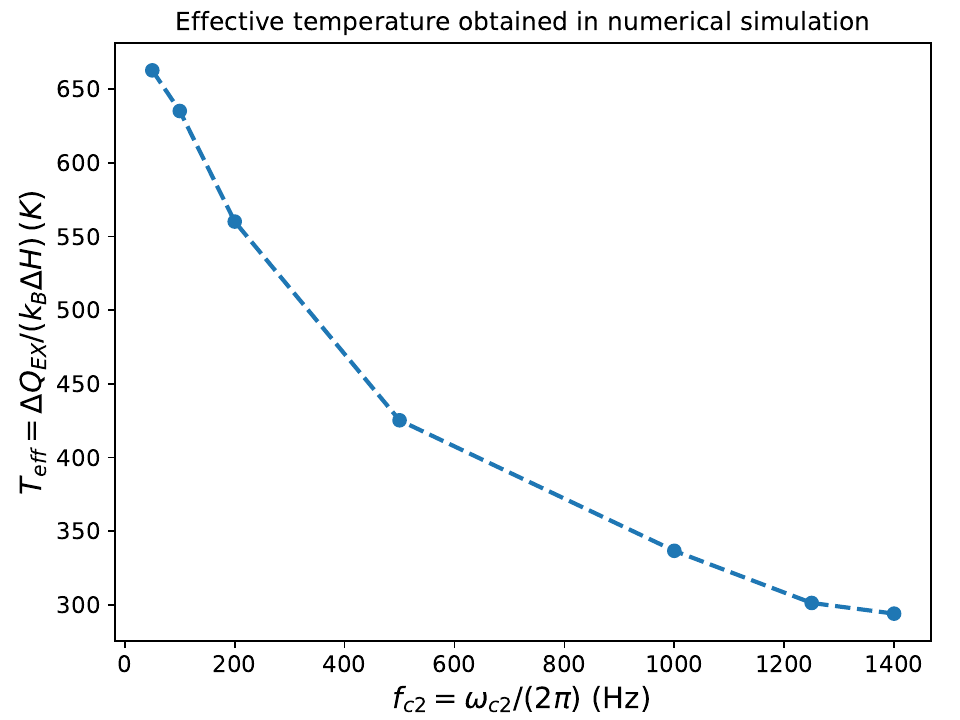}
    \caption{\label{fig:4b}}
\end{subfigure}
\caption{(a) Normalized dissipated heat $\Delta Q_{EX}/(k_\mathrm{B} T_{eq})$ plotted as a function of the spectral entropy difference $\Delta H_S$. The value of $T_{eq}$ is obtained by computing the best linear fit of $\Delta Q_{EX}/k_\mathrm{B}$ as a function of $\Delta H_S$. (b) Effective temperature $T_{eff} = \Delta Q_{EX}/(k_\mathrm{B} \Delta H_S)$ computed for several STEP protocols. As expected, the value is not constant, as the dissipated heat and the spectral entropy are not directly proportional.}
\label{fig:fig4}
\end{figure}

\new{First, we stress that $T_{eq}$ is a quantity only accessible experimentally, and that its value depends on the particular set-up that is used to measure it. In particular, here $T_{eq}$ depends on the properties of the different apparatus used to generate the white noise acting on the trapped particle, such as the dynamic range of the digital-to-analogue card, and the response function of the acousto-optic modulator. Therefore, $T_{eq}$ has no theoretical predicted value, and cannot be easily computed numerically. Moreover, equations using $T_{eq}$ can hardly be seen as general results, since their value is system-dependent. Then,} we note that the authors have plotted $\Delta Q_{EX}/(k_\mathrm{B}T_{eq})$ as a function of $\Delta H_S$ to verify their relation (Figure 4 in~\citep{Goerlich2022}). \new{Here,} we have reproduced this figure with the results of our \new{own} numerical simulations in figure~\ref{fig:4a}. As one can see, \emph{with this particular graphical representation}, it is possible to find that $\Delta Q_{EX} \approx k_\mathrm{B} T_{eq} \Delta H_S$, with $T_{eq} = \SI{622}{\kelvin}$, which is compatible with the experimental values measured by the authors. Moreover, the validity of the linear relation is more likely to be found given that experimental values have inevitable error bars. Yet, it is only an approximation, and not a rigorous result.\bigskip

A more reliable way to verify the relation of direct proportionality~\ref{eq:main_result}, would be to plot the ratio of the dissipated heat and the spectral entropy $T_{eff} = \Delta Q_{EX}/(k_\mathrm{B} \Delta H_S)$, as a function of $f_{c2}$. \new{In particular, this verification does not rely on theoretically unknown quantities, such as $T_{eq}$, or parameters that are difficult to measure experimentally, such as $D_a$.} If the relation of direct proportionality is true, then $T_{eff}$ must be a constant. The values of $T_{eff}$ obtained in our numerical simulations are as shown in figure~\ref{fig:4b}. The quantity is clearly not constant, \new{which is expected since we have already shown that the relation~\ref{eq:main_result} is not verified in the numerical simulations}.

\section{Conclusion}

In conclusion, despite being approximately verified in a nice experimental configuration,  the relation of direct proportionality between the spectral entropy and the dissipated heat $\Delta Q_{EX} \propto \Delta H_S$, is not \new{true in the general case}, and is not rigorously verified even in the framework of the original article that introduced it. \new{We also recall that this relation is not predicted by any theoretical work~\footnote{\new{The authors do not provide any reference to support the validity of this relation, besides their experimental measurements.}}. All this cast serious doubts on the claim that this relation is ``akin to Landauer's principle'', and its interpretation in terms of ``the protocol harvesting information from the colored noise''.} Therefore, \new{we believe that it is very} unlikely that this relation ``could serve as a new tool for the study of non-equilibrium systems in non-trivial baths''.

\bibliography{biblio}

\providecommand{\noopsort}[1]{}\providecommand{\singleletter}[1]{#1}%
\begin{thebibliography}{14}%
\makeatletter
\providecommand \@ifxundefined [1]{%
 \@ifx{#1\undefined}
}%
\providecommand \@ifnum [1]{%
 \ifnum #1\expandafter \@firstoftwo
 \else \expandafter \@secondoftwo
 \fi
}%
\providecommand \@ifx [1]{%
 \ifx #1\expandafter \@firstoftwo
 \else \expandafter \@secondoftwo
 \fi
}%
\providecommand \natexlab [1]{#1}%
\providecommand \enquote  [1]{``#1''}%
\providecommand \bibnamefont  [1]{#1}%
\providecommand \bibfnamefont [1]{#1}%
\providecommand \citenamefont [1]{#1}%
\providecommand \href@noop [0]{\@secondoftwo}%
\providecommand \href [0]{\begingroup \@sanitize@url \@href}%
\providecommand \@href[1]{\@@startlink{#1}\@@href}%
\providecommand \@@href[1]{\endgroup#1\@@endlink}%
\providecommand \@sanitize@url [0]{\catcode `\\12\catcode `\$12\catcode
  `\&12\catcode `\#12\catcode `\^12\catcode `\_12\catcode `\%12\relax}%
\providecommand \@@startlink[1]{}%
\providecommand \@@endlink[0]{}%
\providecommand \url  [0]{\begingroup\@sanitize@url \@url }%
\providecommand \@url [1]{\endgroup\@href {#1}{\urlprefix }}%
\providecommand \urlprefix  [0]{URL }%
\providecommand \Eprint [0]{\href }%
\providecommand \doibase [0]{https://doi.org/}%
\providecommand \selectlanguage [0]{\@gobble}%
\providecommand \bibinfo  [0]{\@secondoftwo}%
\providecommand \bibfield  [0]{\@secondoftwo}%
\providecommand \translation [1]{[#1]}%
\providecommand \BibitemOpen [0]{}%
\providecommand \bibitemStop [0]{}%
\providecommand \bibitemNoStop [0]{.\EOS\space}%
\providecommand \EOS [0]{\spacefactor3000\relax}%
\providecommand \BibitemShut  [1]{\csname bibitem#1\endcsname}%
\let\auto@bib@innerbib\@empty
\bibitem [{\citenamefont {Goerlich}\ \emph {et~al.}(2022)\citenamefont
  {Goerlich}, \citenamefont {Pires}, \citenamefont {Manfredi}, \citenamefont
  {Hervieux},\ and\ \citenamefont {Genet}}]{Goerlich2022}%
  \BibitemOpen
  \bibfield  {author} {\bibinfo {author} {\bibfnamefont {R.}~\bibnamefont
  {Goerlich}}, \bibinfo {author} {\bibfnamefont {L.~B.}\ \bibnamefont {Pires}},
  \bibinfo {author} {\bibfnamefont {G.}~\bibnamefont {Manfredi}}, \bibinfo
  {author} {\bibfnamefont {P.-A.}\ \bibnamefont {Hervieux}},\ and\ \bibinfo
  {author} {\bibfnamefont {C.}~\bibnamefont {Genet}},\ }\bibfield  {title}
  {\bibinfo {title} {Harvesting information to control nonequilibrium states of
  active matter},\ }\href {https://doi.org/10.1103/PhysRevE.106.054617}
  {\bibfield  {journal} {\bibinfo  {journal} {Phys. Rev. E}\ }\textbf {\bibinfo
  {volume} {106}},\ \bibinfo {pages} {054617} (\bibinfo {year}
  {2022})}\BibitemShut {NoStop}%
\bibitem [{\citenamefont {Sekimoto}(1998)}]{Sekimoto1998}%
  \BibitemOpen
  \bibfield  {author} {\bibinfo {author} {\bibfnamefont {K.}~\bibnamefont
  {Sekimoto}},\ }\bibfield  {title} {\bibinfo {title} {{Langevin Equation and
  Thermodynamics}},\ }\href {https://doi.org/10.1143/PTPS.130.17} {\bibfield
  {journal} {\bibinfo  {journal} {Progress of Theoretical Physics Supplement}\
  }\textbf {\bibinfo {volume} {130}},\ \bibinfo {pages} {17} (\bibinfo {year}
  {1998})}\BibitemShut {NoStop}%
\bibitem [{\citenamefont {Seifert}(2012)}]{Seifert2012}%
  \BibitemOpen
  \bibfield  {author} {\bibinfo {author} {\bibfnamefont {U.}~\bibnamefont
  {Seifert}},\ }\bibfield  {title} {\bibinfo {title} {Stochastic
  thermodynamics, fluctuation theorems and molecular machines},\ }\href
  {https://doi.org/10.1088/0034-4885/75/12/126001} {\bibfield  {journal}
  {\bibinfo  {journal} {Reports on Progress in Physics}\ }\textbf {\bibinfo
  {volume} {75}},\ \bibinfo {pages} {126001} (\bibinfo {year}
  {2012})}\BibitemShut {NoStop}%
\bibitem [{\citenamefont {Zaccarelli}\ \emph {et~al.}(2013)\citenamefont
  {Zaccarelli}, \citenamefont {Li}, \citenamefont {Petrosillo},\ and\
  \citenamefont {Zurlini}}]{Zaccarelli2013}%
  \BibitemOpen
  \bibfield  {author} {\bibinfo {author} {\bibfnamefont {N.}~\bibnamefont
  {Zaccarelli}}, \bibinfo {author} {\bibfnamefont {B.-L.}\ \bibnamefont {Li}},
  \bibinfo {author} {\bibfnamefont {I.}~\bibnamefont {Petrosillo}},\ and\
  \bibinfo {author} {\bibfnamefont {G.}~\bibnamefont {Zurlini}},\ }\bibfield
  {title} {\bibinfo {title} {Order and disorder in ecological time-series:
  Introducing normalized spectral entropy},\ }\href
  {https://doi.org/https://doi.org/10.1016/j.ecolind.2011.07.008} {\bibfield
  {journal} {\bibinfo  {journal} {Ecological Indicators}\ }\textbf {\bibinfo
  {volume} {28}},\ \bibinfo {pages} {22} (\bibinfo {year} {2013})},\ \bibinfo
  {note} {10 years Ecological Indicators}\BibitemShut {NoStop}%
\bibitem [{Note1()}]{Note1}%
  \BibitemOpen
  \bibinfo {note} {Both stationarity and ergodicity are required so that the
  ensemble average and time average are equals.}\BibitemShut {Stop}%
\bibitem [{\citenamefont {Wiener}(1930)}]{Wiener1930}%
  \BibitemOpen
  \bibfield  {author} {\bibinfo {author} {\bibfnamefont {N.}~\bibnamefont
  {Wiener}},\ }\bibfield  {title} {\bibinfo {title} {{Generalized harmonic
  analysis}},\ }\href {https://doi.org/10.1007/BF02546511} {\bibfield
  {journal} {\bibinfo  {journal} {Acta Mathematica}\ }\textbf {\bibinfo
  {volume} {55}},\ \bibinfo {pages} {117 } (\bibinfo {year}
  {1930})}\BibitemShut {NoStop}%
\bibitem [{\citenamefont {Khintchine}(1934)}]{Khintchine1934}%
  \BibitemOpen
  \bibfield  {author} {\bibinfo {author} {\bibfnamefont {A.}~\bibnamefont
  {Khintchine}},\ }\bibfield  {title} {\bibinfo {title} {Korrelationstheorie
  der station{\"a}ren stochastischen prozesse},\ }\href@noop {} {\bibfield
  {journal} {\bibinfo  {journal} {Mathematische Annalen}\ }\textbf {\bibinfo
  {volume} {109}},\ \bibinfo {pages} {604} (\bibinfo {year}
  {1934})}\BibitemShut {NoStop}%
\bibitem [{\citenamefont {Mannella}(2002)}]{Mannella2002}%
  \BibitemOpen
  \bibfield  {author} {\bibinfo {author} {\bibfnamefont {R.}~\bibnamefont
  {Mannella}},\ }\bibfield  {title} {\bibinfo {title} {Integration of
  stochastic differential equations on a computer},\ }\href
  {https://doi.org/10.1142/S0129183102004042} {\bibfield  {journal} {\bibinfo
  {journal} {International Journal of Modern Physics C}\ }\textbf {\bibinfo
  {volume} {13}},\ \bibinfo {pages} {1177} (\bibinfo {year}
  {2002})}\BibitemShut {NoStop}%
\bibitem [{Note2()}]{Note2}%
  \BibitemOpen
  \bibinfo {note} {Note that it is also possible to design noises such that
  they do not have the same spectral entropy ($\Delta H_S \protect \neq 0$),
  but produce the same variance of the particle's position ($\Delta Q_{EX} =
  0$).}\BibitemShut {Stop}%
\bibitem [{Note3()}]{Note3}%
  \BibitemOpen
  \bibinfo {note} {Note that the exact values of the parameters are not
  important here, as the numerical results are only presented to ``verify'' the
  analytical calculations, which are expected to hold for any set of
  parameters.}\BibitemShut {Stop}%
\bibitem [{\citenamefont {B\'{e}rut}(2023)}]{codeZenodo}%
  \BibitemOpen
  \bibfield  {author} {\bibinfo {author} {\bibfnamefont {A.}~\bibnamefont
  {B\'{e}rut}},\ }\href {https://doi.org/10.5281/zenodo.7526269} {\bibinfo
  {title} {aberut/browniansimulation1d: v1.5}} (\bibinfo {year}
  {2023})\BibitemShut {NoStop}%
\bibitem [{Git()}]{GithubLink}%
  \BibitemOpen
  \href
  {https://github.com/aberut/BrownianSimulation1D/blob/v1.5/arXiv-2212.06825.ipynb}
  {}\bibinfo {note}
  {\url{https://github.com/aberut/BrownianSimulation1D/blob/v1.5/arXiv-2212.06825.ipynb}}\BibitemShut
  {NoStop}%
\bibitem [{Note4()}]{Note4}%
  \BibitemOpen
  \bibinfo {note} {\textcolor {black}{Note that if the Langevin equation does
  not accurately describe the system, the formulas that are derived from this
  equation are not valid anymore. In particular the definition of the
  stochastic heat~\ref {eq:released_heat}, that is used by the authors in their
  article, does not hold in this case.}}\BibitemShut {Stop}%
\bibitem [{Note5()}]{Note5}%
  \BibitemOpen
  \bibinfo {note} {\textcolor {black}{The authors do not provide any reference
  to support the validity of this relation, besides their experimental
  measurements.}}\BibitemShut {Stop}%
\end{thebibliography}%

\end{document}